\documentclass{bulsao}
\usepackage{graphicx}
\usepackage{latexsym}
\usepackage{amsmath}
\usepackage{amssymb}
\begin{document}

\title{The Most Metal-Poor Quadruple System of Subdwarfs G89-14}

\author{D.~A.~Rastegaev}

\institute{Special Astrophysical Observatory, RAS, Nizhnii Arkhyz,
Karachai-Cherkessian Republic, 357147 Russia}

\offprints{D.~A.~Rastegaev, \\ \email{leda@sao.ru}}

\date{received: October 2, 2009/revised: May 22, 2009}

\titlerunning{The Most Metal-Poor Quadruple System ...}
\authorrunning{Rastegaev}

\abstract{
The system of subdwarfs G89-14 is one of the most metal-poor multiple stars with an
atmospheric metal abundance $\mathrm{[m/H]}=-1.9$. Speckle interferometry at the 6-m BTA telescope
has revealed that G89-14 consists of four components. Measurements of the magnitude difference
between the components and published data have allowed their masses to be estimated:
$M_{A}\approx0.67\ M_{\odot}$, $M_{B}\approx0.24\ M_{\odot}$, $M_{C}\approx0.33\ M_{\odot}$,
and $M_{D}\approx0.22\ M_{\odot}$. The ratio of the orbital periods of the subsystems has been
obtained, 0.52 yr : 3 000 yr : 650 000 yr (1:5769:1250000), indicative of a high degree of
hierarchy of G89-14 and its internal dynamical stability. The calculated Galactic orbital
elements and the low metallicity of the quadruple system suggest that it belongs to the Galactic halo.
}

\maketitle

\section{INTRODUCTION}
Multiple star systems (with $\geqslant3$ components) are currently believed to be a natural result of
star formation (Kroupa et al. 2003; Delgado-Donate et al. 2003). At least $8\%$ of the solar-type
stars consist of three or more components (Tokovinin 2008). Since multiple systems are
characterized by additional parameters compared to single and double stars (the angles
between the orbital planes of various subsystems, eccentricity ratios, etc.), their study
allows more stringent constraints to be imposed on the star formation process. Star
formation models must predict the properties of such objects, while the theory of
dynamical evolution must predict their survivability under both internal (tidal interactions)
and external (the influence of the Galaxy's nonuniform gravitational field) effects.

Investigating metal-poor multiple stars $(\mathrm{[Fe/H]}<-1)$ is important in understanding the
stability of star systems, because the stars with a low atmospheric metal abundance were
generally formed at the epoch when our Galaxy was born, i.e., more than 10 Gyr ago. Studying
multiple systems that have survived over such a long period allows one not only to test various
criteria for dynamical stability of multiple stars but also to impose constraints on the mass
distribution in the Galaxy.

In 2006 and 2007, we conducted a speckle interferometric survey of
223 nearby F, G, and early-K subdwarfs ($\leq250$ pc) with low metallicities $(\mathrm{[Fe/H]}<-1)$
and large proper motions ($\mu \geq 0.26''/$yr ) at the 6-m BTA telescope of the Special Astrophysical
Observatory, Russian Academy of Sciences (Rastegaev et al. 2007, 2008; Rastegaev 2009). The goal of
the survey of Population II stars, which was carried out with the diffraction-limited angular
resolution of the 6-m telescope ($0.023 ''$ at 550 nm), was to expand the database of double and
multiple old stars and to determine the orbital parameters and properties of their components. One
of the results of this survey was the discovery of the quadruple system of subdwarfs G89-14
(HIP 35756).

In this paper, we present the fundamental parameters of this system obtained from our
observations and published data.

\begin{table*}
\begin{center}
\caption{Results of speckle measurements for G89-14}
\label{obs}
\bigskip
\begin{tabular}{ c | c | c | c | c | c | c | c }
\hline
$\rho\ ('')$ & $\sigma_{\rho}$ &$\Theta\ (^{\circ})$ & $\sigma_{\Theta}$ &$\Delta m$ & $\sigma_{\Delta m}$ & $\lambda/\Delta\lambda$, nm & Epoch \\
      &   &   &   &    &  &  &  \\
\hline
0.979 & 0.009 & 0.8 & 0.4 & 4.1 & 0.4  & 800/100  & 2006.94455 \\
0.982 & 0.005 & 0.8 & 0.4 & 4.3 & 0.1  & 800/100  & 2007.24040 \\
\hline
\multicolumn{6}{c|}{Unresolved}      & 550/20   & 2007.24040 \\
\hline
\end{tabular}
\end{center}
\end{table*}

\vspace*{0.3cm}
\section{OBSERVATIONS}
The speckle observations of G89-14 were performed at the 6-m BTA telescope of the Special Astrophysical
Observatory, Russian Academy of Sciences, in December 2006 and March 2007 (Rastegaev et al. 2007, 2008).
In the observations, we used a system (Maksimov et al. 2009) based on a 512$\times$512-pixel EMCCD
(Electron Multiplying CCD) with a high quantum effciency and linearity, which allowed objects with
a magnitude difference between the components $\Delta m \lesssim 5^{m}$ to be discovered with the
diffraction-limited resolution of the 6-m telescope. The size of the detector field, 4$''$ made it possible
to detect the secondary components at angular distances as large as 3$''$ from the primary star.

In December, under good weather conditions (seeing $<1''$), we accumulated 500 20-ms exposures for G89-14 through
a 800/100 nm filter (the first and second numbers give the central wavelength of the filter passband and the
passband FWHM, respectively). In March, the weather conditions were not optimal for speckle observations
($\approx 3''$). We obtained 2000 speckle images with 20-ms exposures through each of the two filters,
550/20 and 800/100 nm.

We calibrated the measurements based on the so-called ``standard'' pairs --- binary
systems with well known separations between the components and position angles. The technique for determining
the relative positions and magnitude differences of the components of the objects under study from speckle
interferograms averaged over a series of power spectra was described by Balega et al. (2002). The accuracy
achieved with this technique is $0.02^m$ for the magnitude difference, 1 mas for the angular separation,
and $0.1^{\circ}$ for the position angle.

The results of our measurements for the speckle interferometric subsystem of G89-14
are presented in Table 1. The March observations through the 550/20 filter did not reveal
the speckle interferometric component at a distance of $0.98''$ from the known spectroscopic pair
(see below), because its magnitude in this spectral range was fainter than that of the SB1 system by
more than $5^m$. At the same time, the observations through the 800/100 filter in March 2007 confirmed,
within the error limits, the results of our interferometric observations in December 2006.

\vspace*{0.3cm}
\section{G89-14: A SYSTEM OF FOUR SUBDWARFS}

\begin{figure}
\begin{center}
\includegraphics[width=80mm,height=80mm]{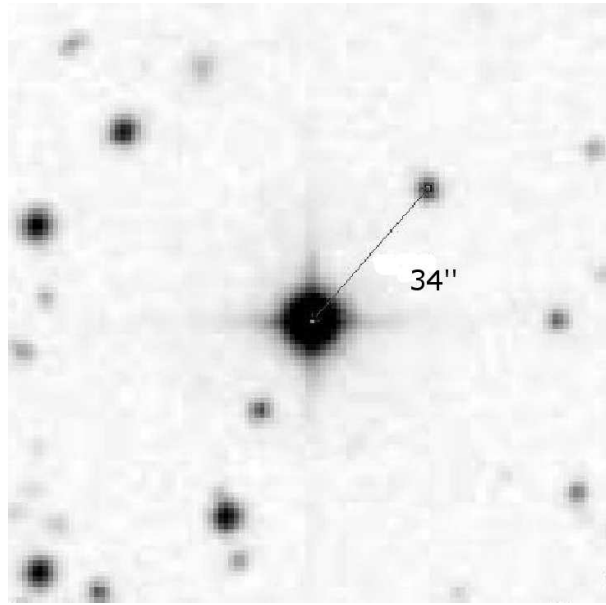}
\caption{\rm Image of G89-14 from the DSS2 red survey: the triple subsystem consisting of the spectroscopic pair and
the speckle interferometric component is at the center; the component with a common proper motion is located
$34''$ to the northwest. In the figure taken from the Aladin database, the north is at the top and the east
is to the left.}\label{G89-14}
\end{center}
\end{figure}

Initially, G89-14 (TYC 763-801-1; HIP 35756; WDS 07224+0854; NLTT 17770) with coordinates
$07^h22^m31\fs5$ $+08^\circ49' 13.0''$  (J2000.0) was known as an SB1-type spectroscopic binary
system with a period of about 190 days (Latham et al. 1988, 2002). Subsequently, Allen et al. (2000)
found a companion to this pair with a common proper motion located at an angular distance of $34''$.
Unfortunately, these authors did not give the position angle of this companion. In the Aladin database,
there is only one star at a distance of $34''$ that we took as the component with a common proper motion
(Fig. 1). Allen et al. (2000) refuted the physical connection of G89-14 with G89-13 (HIP 35750), because,
having a common proper motion, the parallax and radial velocity of G89-13 differed significantly from
those of G89-14. In 2006, using the BTA telescope, we discovered a speckle interferometric component
at an angular distance of $\approx1''$ from the spectroscopic binary (Rastegaev et al. 2007, 2008).
According to the Hippacros data, the heliocentric distance of G89-14 is $\approx170$ pc (ESA 1997).
The distance determined by Carney et al. (1994) from the photometric parallax is $94$ pc and is most
likely an underestimate, because it disregards the presence of additional components. At a distance
of 170 pc from the Sun, the probability that the speckle interferometric companion at an angular
distance of $0.98''$ and with a magnitude difference of about $4^m$ is an optical projection is
insignificant. Hence, we concluded that the discovered companion and the system G89-14 are physically
connected. Their gravitational connection is also confirmed by the repeated interferometric observations
in March 2007. If the component were an optical projection, then, given the large proper motion of G89-14,
$\mu\approx0.3''/$yr (Carney et al. 1994; ESA 1997), the relative position of the component would change
by $\approx0.1''$ in the time between our observations (from December to March), which exceeds the errors
of our speckle measurements. Thus, four gravitationally bound components of the system are known at present
(Fig. 2). The metallicity of G89-14 is $\mathrm{[m/H]}=-1.9$ (Carney et al. 1994). Analysis of the literature has
shown that none of the quadruple stars known to date has such a low metal abundance.

\vspace*{0.3cm}
\section{ESTIMATES OF THE COMPONENT MASSES}
\begin{figure}
\begin{center}
\includegraphics[width=60mm,height=60mm]{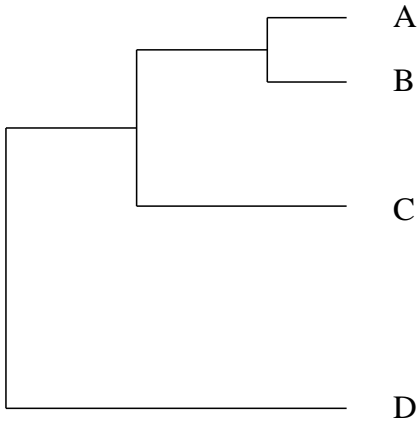}
\caption{\rm Schematic representation of the hierarchy levels for G89-14: A--B is the spectroscopic
subsystem, AB--C is the speckle interferometric subsystem, and ABC--D is the subsystem with a common proper motion.
}\label{hierarchy}
\end{center}
\end{figure}
Accurate stellar masses can be determined only in binary and multiple systems by constructing their orbits.
In the remaining cases, only approximate estimates based on various theoretical and empirical relations
(e.g., the mass-luminosity relation) are possible. All orbital parameters are known for none of the
subsystems of G89-14. A spectroscopic orbit was obtained for the shortest-period subsystem with a period
of $190.49\pm0.29$ days (Latham et al. 2002). However, in this case, the inclination $i$ of the orbital
plane to the plane of the sky remains unknown.

Using evolutionary tracks, spectroscopic (Carney et al. 1994) and photometric (Allen et al. 2000;
Monet et al. 2003) data, and our speckle observations, we estimated the
masses of the components in the quadruple system under consideration.

Based on the tracks from
Vandenberg and Bell (1985) and the spectroscopically measured metallicity and temperature of the primary
component (5750 K) of the spectroscopic subsystem of G89-14, Carney et al. (1994) estimated its mass to be
$0.67\ M_{\odot}$. The mass function that they obtained from spectroscopic observations,
\begin{equation}
\label{mass_function}
 \frac{M_{B}^{3}\sin^{3}i}{(M_{A}+M_{B})^2} = 0.0173,
\end{equation}
allows the mass of the secondary spectroscopic component to be determined to within a factor of $\sin^{3}i$.
Substituting $M_{A}=0.67\ M_{\odot}$ into Eq. (1) and setting $\sin^{3}i=1$, we obtain the lower mass limit
$M_{B}\geq0.24\ M_{\odot}$.

The mass of the speckle interferometric companion was estimated from the total
magnitude of the triple subsystem (the SB1 pair and the interferometric component), $10.0^{m}$ in the $I$
band (Monet et al. 2003), and from the magnitude difference between the spectroscopic binary and the
companion we discovered (Table 1). For our calculations, we took $\Delta m = 4.2$ in the 800/100 filter,
which roughly corresponds to the $I$ band. The apparent magnitude of the speckle interferometric component
was $14.22^{m}$. Based on the absolute magnitude of the companion ($M_{I}=8.07^m$) and using the evolutionary
tracks for $\mathrm{[Fe/H]}=-2.0$ from Baraffe et al. (1997), we estimated the mass of the interferometric
component to be $\approx0.33\ M_{\odot}$.

Allen et al. (2000) gave the absolute magnitude of the most distant component $M_{V}=10.4^{m}$,
which, according to the evolutionary tracks from Baraffe et al. (1997), corresponds to a mass of
$0.22 M_{\odot}$ for the metallicity $\mathrm{[Fe/H]}=-2.0$. Judging by the mass function from
Chabrier (2003) derived for the Galactic halo stars, this mass is a mean for the halo field stars.
Note that the distance to the star used by Allen et al. (2000) is slightly larger, 180 pc instead
of 170 pc that we adopted. We did not apply a correction to the mass estimate for this component
due to the difference between the assumed heliocentric distances.

Thus, denoting the primary component of the spectroscopic pair, its closest companion,
the speckle interferometric component, and the most distant component
with a common proper motion by A, B, C, and D, respectively (Fig. 2), we have
$M_{A}\approx0.67\ M_{\odot}$, $M_{B}\approx0.24\ M_{\odot}$, $M_{C}\approx0.33\ M_{\odot}$,
$M_{D}\approx0.22\ M_{\odot}$.

\vspace*{0.3cm}
\section{ORBITAL PERIODS OF THE SUBSYSTEMS AND DYNAMICAL STABILITY}
Using the above mass estimates and Kepler's generalized third law, we calculated the unknown periods
of two subsystems of G89-14. The semimajor axis appearing in Kepler's law was determined for each
subsystem from an empirical relation between the projected component separation and the orbital
semimajor axis. Given the parallax of the system $\pi$ and the projected separation between the
components $\rho$, the expected semimajor axis can be calculated from the formula (Allen et al. 2000)
\begin{equation}
 \left\langle a \right\rangle =10^{\mathrm{log}\frac{\rho}{\pi} + 0.146} ,
\end{equation}
where $\left\langle a \right\rangle$ is in astronomical units, $\rho$ and $\pi$ are in arcseconds.

Our calculations show that the speckle interferometric component and the SB1 pair make one revolution around
the common center of mass approximately in 3000 yr ($P_{AB-C}$), while the revolution period of the most
distant component and the triple subsystem is about 650000 yr ($P_{ABC-D}$). For the outer subsystem, we
took the expected semimajor axis $\left\langle a \right\rangle = 8565$ AU from Allen et al. (2000). The
ratio of the subsystem periods, 1 : 5769 : 1250000, is indicative of a high degree of hierarchy of G89-14
and, hence, its internal dynamical stability. Nevertheless, the outer subsystem has a low binding energy
and is subjected to the destructive effect from giant molecular clouds and the stellar component of our
Galaxy (Weinberg et al. 1987).

\vspace*{0.3cm}
\section{GALACTIC ORBIT}
\begin{table*}
\begin{center}
\caption{Initial conditions for constructing the Galactic orbit of G89-14}
\label{initial}
\bigskip
\begin{tabular}{ c | c | c | c | c | c | c | c}
\hline
$v_{rad}$, km/s & $\mu$, $''$/yr & $D$, pc & $\widetilde{\omega}$, kpc & $z$, kpc & $\Phi$, km/s & $W$, km/s & $h$, km$\cdot$kpc/s \\
\hline
$-36.9$         & 0.31            & 170     & $8.15$                    & $0.03$   & $-165$       & $13$      & $194$ \\
\hline
\end{tabular}
\end{center}
\end{table*}

To ascertain which subsystem of our Galaxy (halo, thick or thin disk) G89-14 belongs to, apart  from the
atmospheric metal abundance, it is also necessary to know the pattern of motion of the object in the Galactic
gravitational field. The most complete description of the star's dynamical properties is the construction of
its Galactic orbit.

To construct the Galactic orbit, we must find a solution to the equations of motion in
the gravitational potential of the Galaxy. For an axisymmetric potential, the specific angular momentum is
a conserved quantity and the motion can be described in a moving meridional plane. In cylindrical coordinates,
the motion of the star in the meridional plane is described by a system of second-order differential equations:
\begin{equation}
 \label{motion_w}
 \frac{d^2 \widetilde{\omega}}{dt^2}=-\frac{\partial \phi}{\partial \widetilde{\omega}}+\frac{h^2}{\widetilde{\omega}^3},
\end{equation}
\begin{equation}
 \label{motion_z}
\frac{d^2 z}{dt^2}=-\frac{\partial \phi}{\partial z}.
\end{equation}
At the same time, the motion of the meridional plane itself is specified by the equation
\begin{equation}
 \label{motion_eta}
 \frac{d \theta}{dt}=\frac{h}{\widetilde{\omega}^2},
\end{equation}
where $\phi$ is the gravitational potential of our Galaxy, $h=\Theta \cdot \widetilde{\omega}$ is the specific
angular momentum of the body, $\widetilde{\omega}=\sqrt{X^2+Y^2}$, $\theta=\mathrm{arctg}(Y/X)$ and $z=Z$ are
the cylindrical Galactic coordinates, and  $\Theta$ is the projection of the space velocity vector corresponding
to coordinate $\theta$; $X$, $Y$ and $Z$ are the coordinates of the object in the rectangular coordinate system
with the origin at the Galactic center and with the $XY$ plane coincident with the Galactic plane. The second
term on the right-hand side of Eq. (3) is the Coriolis acceleration acting on the body in the rotating frame
of reference.

The initial conditions in solving the equations of motion (3)--(5) are the position and velocity
of the star in the Galaxy. To specify the initial conditions, we must know six observable quantities: the
object's coordinates ($l$ and $b$), proper motions in $\alpha$ ($\mu_{\alpha}$, $''$/yr ) and $\delta$
($\mu_{\delta}$, $''$/yr), parallax ($\pi$, $''$), and radial velocity ($v_{rad}$, km/s). Table 2 presents
the input data for integrating the Galactic orbit of G89-14. The projection of the space velocity vector
onto $\widetilde{\omega}$ is denoted by $\Phi$. In our calculations, we assumed the distance from the Galactic
center to the Sun to be $\widetilde{\omega}_{\odot}=8$ kpc and the solar motion relative to the Local Standard
of Rest to be $(U_{\odot}, V_{\odot}, W_{\odot})$ = ($-10.0$, $5.0$, $7.0$ km/s) (Dehnen and Binney 1998).
Using formulas from Johnson and Soderblom (1987), we passed from the observable quantities
(ESA 1997; Latham et al. 2002) to the projections of the space velocities of G89-14 in the rectangular
coordinate system, $(U, V, W)$ = ($-165$, $-195$, $13$ km/s).

To calculate the parameters of the body's Galactic orbit, it will
suffice to solve two equations of motion, (3) and (4), numerically. For the solution,
we used the seventh-order Runge--Kutta--Nystr$\mathrm{\ddot{o}}$m method with a variable step
(see, e.g., Fehlberg 1972). The parameters of the derived Galactic orbit of G89-14 are presented
in Table 3. As the gravitational potential of our Galaxy, we chose a three-component model described by
Allen and Santillan (1991). Figure 3 shows the orbit of G89-14. Over $10^{10}$ yr, the system has crossed
the Galactic plane $\sim100$ times. G89-14 may have had a larger number of components that were lost
through the tidal effect from the Galaxy, because part of the time the system moved in the Galactic plane,
where this effect is maximal. We see from the figure that during its multiple revolutions, the star has
closely approached the Galactic center ($\widetilde{\omega}_{min}=0.3$ kpc). The orbit is clearly
chaotic and has a high eccentricity, which was determined from the formula
\begin{displaymath}
 e=\frac{R_{apo}-R_{peri}}{R_{apo}+R_{peri}},
\end{displaymath}
where $R=\sqrt{\widetilde{\omega}^2+z^2}$; $R_{apo}$ and $R_{peri}$ are the maximum and minimum distance
of the star from the Galactic center. The orbital parameters, along with the low metallicity, allow us
to confidently classify the object under study as belonging to the halo population.

Carney et al. (1994) also obtained the parameters of the Galactic orbit for G89-14, suggesting that the
star most likely belongs to the thin or thick disk. Given the low atmospheric metal abundance in the star,
its belonging to the Galactic disk is less plausible. The main error in their calculations was the
underestimated heliocentric distance (94 pc) determined by Carney et al. from the photometric parallax.

\begin{table*}
\begin{center}
\caption{Parameters of the Galactic orbit for G89-14}
\label{result}
\bigskip
\begin{tabular}{ c | c | c | c | c }
\hline
$\widetilde{\omega}_{min}$ & $\widetilde{\omega}_{max}$ & $z_{min}$ & $z_{max}$ & $e$ \\
\hline
$0.3$ kpc & $10.1$ kpc & $-6.5$ kpc & $6.4$ kpc & $0.93$  \\
\hline
\end{tabular}
\end{center}
\end{table*}

\begin{figure}
\begin{center}
\includegraphics[width=90mm,height=90mm]{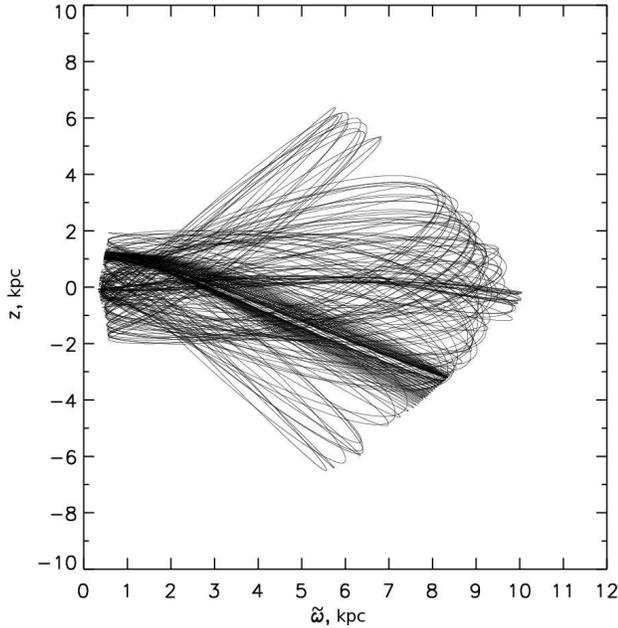}
\caption{\rm Galactic orbit of G89-14.}\label{gal_orbit}
\end{center}
\end{figure}

\vspace*{0.3cm}
\section{CONCLUSIONS}
Our speckle observations at the 6-m BTA telescope revealed that G89-14 consists of four subdwarfs. A review
of the literature showed that G89-14 with $\mathrm{[m/H]}=-1.9$ is the most metal-poor quadruple system
known to date. This made it an interesting object for a more detailed study.

Based on evolutionary tracks
(Baraffe et al., 1997), available spectroscopic (Carney et al. 1994) and photometric (Allen et al. 2000;
Monet et al. 2003) data, and our speckle interferometry, we estimated the masses of the components of
G89-14: $M_{A}\approx0.67\ M_{\odot}$, $M_{B}\approx0.24\ M_{\odot}$, $M_{C}\approx0.33\ M_{\odot}$,
$M_{D}\approx0.22\ M_{\odot}$. The ratio of the orbital periods of the subsystems
$P_{AB} : P_{AB-C} : P_{ABC-D} =$ 0.52 yr : 3000 yr : 650000 yr (1:5769:1,250,000) is indicative
of a high degree of hierarchy of G89-14 and, consequently, its high internal dynamical stability. Using
the three-component model of the gravitational potential by Allen and Santillan (1991), we constructed
the Galactic orbit of the quadruple star under consideration (Fig. 3). Analysis of the Galactic orbital
elements and the low metallicity of the system suggest that G89-14 belongs to the Galactic halo. Further
observations of G89-14 by various methods are needed to improve the orbital elements of the subsystems
and the fundamental parameters of the system's components.

\begin{acknowledgements}
I wish to thank the staff of the group of high angular resolution astronomy methods from the Special
Astrophysical Observatory of the Russian Academy of Sciences for help with the observations and separately
Yu.Yu. Balega and E.V. Malogolovets for valuable remarks when writing this paper as well as A.F. Valeev
for advice on code optimization. The Galactic orbit was calculated using the IDL and MAPLE software packages.
This work was performed using the SIMBAD database and was supported by the Russian Foundation for Basic
Research (project no. 04-02-17563) and the Program of the Division of Physical Sciences of the Russian
Academy of Sciences.

Translated by G. Rudnitskii.
\end{acknowledgements}

\end{document}